\documentclass[journal=jpc]{achemso}

\usepackage{titlesec}
\usepackage{indentfirst}
\usepackage{hyperref}
\usepackage{siunitx}
\usepackage{amsmath}
\usepackage[normalem]{ulem}
\usepackage{soul}
\usepackage{adjustbox}
\usepackage[table, svgnames]{xcolor}
\title{Island Sliding Barriers: A first-principles metric for determining remote epitaxy viability}

\author{Quinn T. Campbell}
\affiliation{Center for Computing Research, Sandia National Laboratories, Albuquerque, NM, USA}
\email{qcampbe@sandia.gov}
\author{Manny Xavier de Jesus Lopez}
\author{Anthony Rice}
\author{Timothy J. Ruggles}
\author{Taisuke Ohta}
\author{Caitlin McCowan}
\author{Sadhvikas Addamane}
\author{Scott W. Schmucker}
\author{Justine Koepke}
\affiliation{Sandia National Laboratories, Albuquerque, NM, USA}
\date{\today}

\begin{document}


\abstract{
Remote epitaxy, where a 2D van der Waals material (usually graphene) is inserted on top of the substrate before film epitaxy,  has emerged as a promising path for growing electronics with lower defect rates and less stringent lattice matching requirements. 
The exact mechanism behind remote epitaxy has not been definitively shown, however, and it is not obvious when examining a new substrate-film pair whether they would be compatible with the remote epitaxy process. 
In this paper, we use first principles calculations to test several different mechanisms for determining whether a given substrate-film pair will successfully be grown with remote epitaxy.
We find that previously calculated metrics such as electrostatic potential do not hold sufficient explanatory power. 
We find that the sliding barrier of small islands on the surface when the atomic positions are allowed to optimize provides the most rigorous criteria for whether a given substrate-film pair is remote epitaxy active. 
This indicates that remote epitaxy is likely a phenomenon related to the kinetics and ease of island migration on the graphene surface.
}

\maketitle

\section{Introduction}\label{sec:intro}
Why does remote epitaxy work? % too informal, but that's also why I like it
This has been the key question ever since the process was introduced by Kim \textit{et al.}\cite{kim2017remote}.
In remote epitaxy, a 2D van der Waals layer of material (usually graphene) is placed over the substrate and then a growth process is initiated similar to traditional epitaxy.
Remote Epitaxy (RE) has been shown to have numerous potential advantages over both traditional epitaxy and van-der-Waals epitaxy.
The interaction of the graphene with the underlying substrate can lead to less stringent lattice matching requirements.
Furthermore, the epilayer has typically been shown to be single domain, with reduced misfit dislocations and other defects, and with the ability to be rapidly released and transferred \cite{chang2023remote,kim2023high}. 
This combination of traits opens up exciting semiconductor growth possibilities and could be crucial for the development of next generation opto- and power electronics \cite{kim2022remote,park2024remote}.

While RE provides new semiconductor growth opportunities, the exact physical/chemical mechanism that enables RE remains under debate.
The dominant explanation put forward so far by Kong \textit{et al.} \cite{kong2018polarity} is that the polarity of the substrate dominates whether it will be active for RE.
Under this theory, substrates with stronger polarity, such as GaN or LiF will be able to transmit their underlying lattice through more layers of graphene. 
This hypothesis is backed up by examining both the electrostatic potential from Density Functional Theory (DFT) calculations and Electron Back Scattering diffraction (EBSD) characterization of grown homo-epitaxial layers, finding single domain growth extends across more graphene layers for more polar materials.
While this proposed chemical mechanism is tantalizing, it is not fully explanatory.
Does polarity apply in cases where the substrate and film differ? 
If so, how exactly would the interaction be moderated between two materials with different levels of polarity?
Furthermore, this theory assumes a perfect graphene layer is formed which does not affect the underlying substrate and is essentially transparent to the underlying substrate's bonding patterns. 

Recent work has begun to question a number of key aspects around how perfect and transparent the graphene on top of the substrate actually is. 
A recent analytical model showed that even one layer of graphene (treated as homogenous electron gas with the carrier density of graphene) would lead to multiple orders of magnitude weakening of the underlying bonding potential \cite{kawasaki2025analytical}.
Under the same model, the dominant factor influencing how much of the remote epitaxy potential goes through is not the polarity of the substrate, but how extended its bonding orbitals are.
Furthermore, what if the graphene undergoes a reconstruction with the underlying substrate?\cite{laduca2025transparent}
Presumably the graphene can no longer be considered to not be interacting with the substrate and its bonding with the film.
And finally there remains the ever-present possibility that the graphene layers have significant pinholes, which would lead to direct epitaxy in significant regions of the graphene and account for much of the observed RE characteristics even if these pinholes are atomic in size \cite{du2022controlling}. 
\begin{figure}
\includegraphics[width=0.7\textwidth]{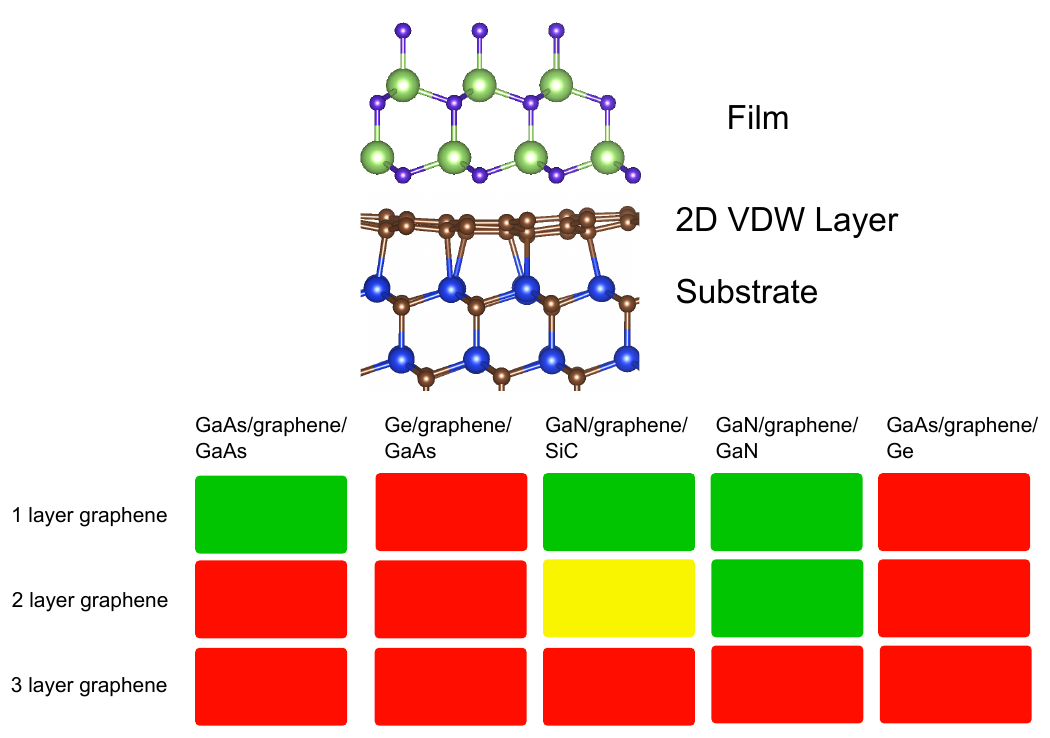}
\caption{ A schematic of a typical remote epitaxial (RE) scheme, where a substrate is covered by a 2D van der Waals (vDW) layer, which is then used to grow high quality, single crystal films that would not previously be available for the substrate.
We then present a schematic of which systems have been found to be viable with RE, with green boxes representing that experimental evidence has been shown with single crystalline layers from these setups, red indicating that only multicrystalline, vDW like layers have been grown, and yellow indicating that the evidence is ambiguous. 
The data for GaAs/graphene/GaAs comes from Ref.~\cite{kim2017remote}, Ge/graphene/GaAs and GaAs/graphene/Ge from Ref.~\cite{chang2023remote}, GaN/graphene/SiC from Ref.~\cite{journot2019remote,qiao2021graphene,chang2023remote}, and GaN/graphene/GaN from Ref.~\cite{kong2018polarity}.
\label{fig:intro}}
\end{figure}

Fundamentally, it still remains unclear for a novel pair of substrate and film if these materials would be compatible with a RE growth process, and, if so, how many of layers of graphene would the process work with. 
In Fig.~\ref{fig:intro} we schematically show a set of substrate and film pairings that have been reported in the literature to be active as well as inactive for RE as well as how many layers of graphene this is true for. 
Even substrates that have shown RE can interact poorly with different films as demonstrated by GaAs on GaAs succeeding, while Ge on GaAs does not.
While there has been significant discussion on complicating factors in the experimental preparation of these results (density of pinholes often being a key unknown), for the purposes of this paper, we will accept the experimental results as providing a ground truth for the viability of RE in different systems.
To advance next generation electronic materials growth, the field needs an (ideally first-principles) metric which could reliably predict whether a given substrate and film pairing will work with RE and with how many graphene layers. 

In this manuscript, we use first-principles calculations of a variety of different substrate/film combinations to determine a metric for predicting to RE.
We examine previous metrics and mechanisms such as charge density extension and electrostatic potential decay, finding them insufficient to explain experimental results. 
We also examine the properties of individual atoms above the substrate graphene, finding them insufficient to explain RE.
Finally, we determine the sliding barrier of film islands on the graphene/substrate surface as the most reliable indicator of RE viability.
These results imply that the RE growth process is dominated by kinetics and future work should focus on developing more detailed understandings of the initial island growth and agglomeration.

\section{Methodology}\label{sec:methods}

For each system of interest, we created slab models where the bottom layers of the substrate were terminated with a passivating hydrogen atom, and the top of the substrate was exposed to graphene layers and then film atoms or islands as specified in the results section.
All the exact structures and input files used for these analysis can be found in the supplementary material for this work. 
For each structure, the atomic geometry is then optimized using DFT calculation.

DFT calculations are done using the Quantum {\sc espresso} package~\cite{giannozzi2009quantum}.
We use norm-conserving pseudopotentials from the PseudoDojo repository~\cite{van2018pseudodojo} and the Perdew-Burke-Ernzerhof exchange-correlation functional~\cite{perdew1996generalized}.
We use kinetic energy cutoffs of 50 Ry and 400 Ry for the plane wave basis sets used to describe the Kohn-Sham orbitals and charge density, respectively.
We use a 2$\times$2$\times$1 Monkhorst-Pack grid~\cite{monkhorst1976special} to sample the Brillioun zone in our calculations.
Equilibrium structures are found by allowing the forces to relax below 0.05 eV/\AA.

For electrostatic potential calculations, the planar electrostatic potential is calculated 3 \AA\ above the graphene layer in question. 
For cases such as the SiC $6\sqrt{3}$ reconstruction where the graphene layer has vertical variation, we calculate the planar electrostatic potential 3 \AA\ above the average position of the carbon atoms in the graphene. 

For the single atom adsorption energies, we place a single atom of the relevant species 3 \AA\ above the graphene surface. 
For the resulting atomic optimization, we then allow the film atom to optimize its location only in the vertical $z$ direction.
This preserves the registry of the adsorption energy with the underlying substrate, avoiding the possibility of placing atoms at different horizontal locations across the substrate and having them all optimize to the same location.
The adsorption energy $E_a$ is calculated as 
\begin{equation}
    E_a = E_{slab/atom} - E_{slab} - E_{atom},
\end{equation}
where $E_{slab/atom}$, $E_{slab}$, and $E_{atom}$ represent the DFT calculated energy of the slab and atom optimized together, the slab on its own, and the atom on its own respectively. 

Similarly, we calculate the bonding energy $E_b$ of an island above a substrate as 
\begin{equation}
    E_b = E_{slab/island} - E_{slab} - E_{island},
\end{equation}
where $E_{slab/island}$ and $E_{island}$ are the DFT energy of the island on top of a slab and an isolated island, respectively. 
We calculate $E_b$ for the slab at a spacing of every 1 \AA\ along a straight line along the slab. 
We then calculate the sliding barrier $E_{sliding}$ as 
\begin{equation}
    E_{sliding} = max(E_b) - min(E_b)
\end{equation}
for a given system. 
This may miss saddle points along the system that occur in between the every 1 \AA\ spacing of our model and does not give as robust results as a full nudged elastic band calculation. 
These therefore serve as effective lower bounds on the sliding barrier for a given island/substrate system.
Due to the size of the islands being $>1$ \AA, however, it is likely that these lower bounds are reasonably converged. 

\section{Results}\label{sec:results}
We present our search for different metrics that define RE as a series of lessons on how modeling at different levels of fidelity and detail can impact results.

\subsection{Graphene Matters}
\begin{figure*}
\includegraphics[width=\textwidth]{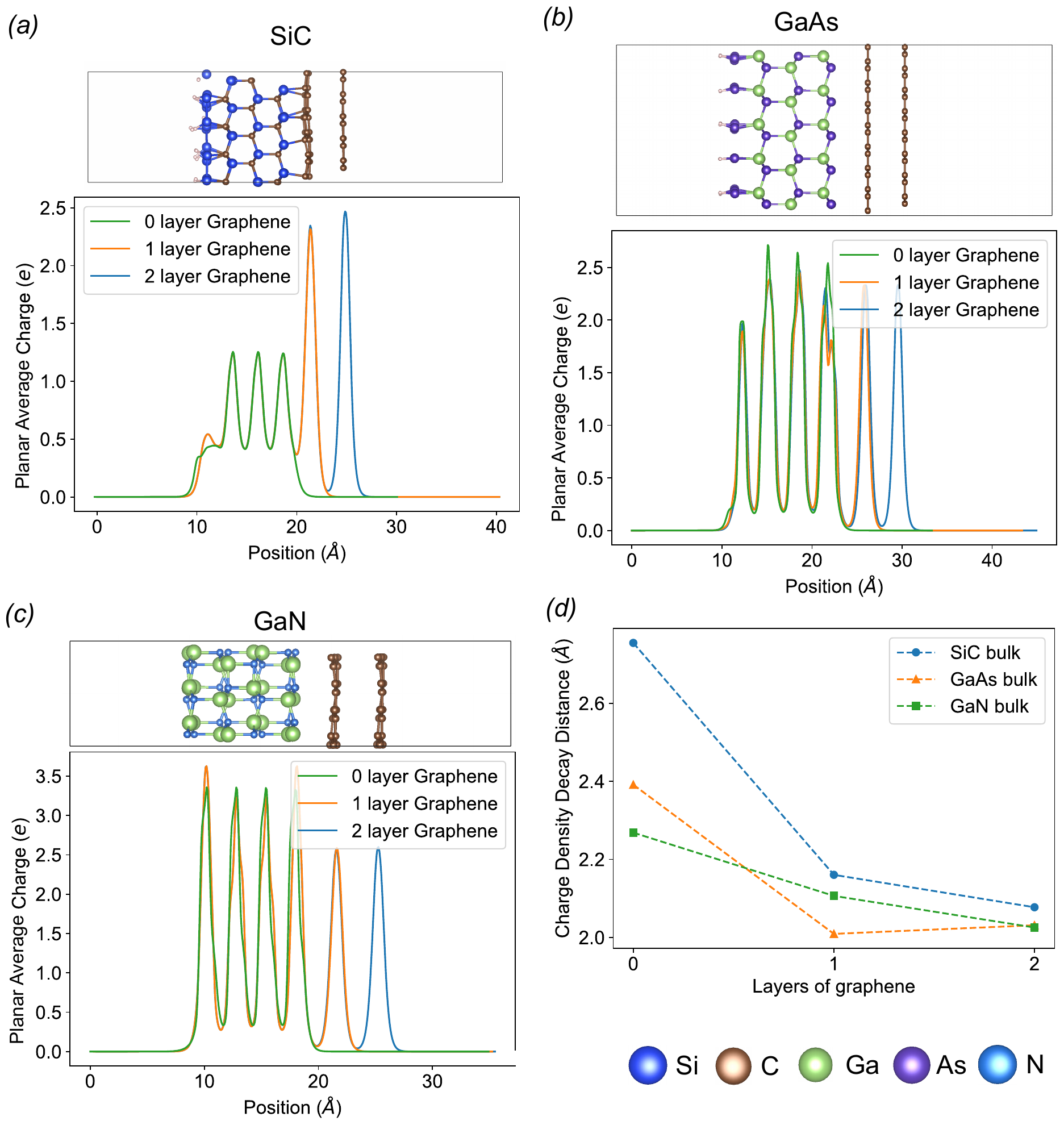}
\caption{ The planar average charge moving toward the surface for (a) SiC, (b) GaAs, and (c) GaN. We then examine (d) how the decay distance of each material changes as a function of the number of graphene layers.  
\label{fig:chg-density}}
\end{figure*}

We begin by examining the spread of charge densities as a potential metric for RE viability. 
We mimic this analysis from early results from Kim \textit{et al.} \cite{kim2017remote} which was further refined in similar charge density analysis in other works \cite{jeong2018remote,jiang2019carrier,jeong2020remote}.
First, we generate slabs of SiC, GaAs, and GaN as shown in Fig.~\ref{fig:chg-density}. 
Then we measure how far the charge density of these materials extends into the vacuum. 
The hypothesis would be that substrates with stronger RE capability have charge densities that extend further into the vacuum, potentially allowing them to bond ``through'' graphene layers. 
However, as shown in Fig.~\ref{fig:chg-density}d, this hypothesis does not align with the observed decay distances. 
We define the charge density decay distance $z_{decay}$ as the distance from the peak of the surface charge $z_{peak}$ until the planar average charge goes below a threshold of 0.001 $e$ $z_{thr.}$, i.e. $z_{decay}= z_{thr.} - z_{decay}$.
The decay distance for all of these materials is in the range of 2.0 - 2.7 \AA.
Notably, this charge decay distance is less than the average distance from the surface to a graphene layer ($\approx$ 3 \AA).
Any charge transfer ``through'' the graphene would then have to take advantage of covalent bonding with the graphene.

Furthermore, it is not clear that the charge decay distances meaningfully correlate with RE in any way. 
While the charge decay distance does generally decrease as the number of graphene layers is increased, this is not a uniform trend: for GaAs, the charge decay distance is less at 1 layer than at 2 layers. 
Would this imply that a GaAs system with 2 layers of graphene would be better able to sustain remote epitaxy? 
This contradicts experimental EBSD evidence \cite{kim2017remote}.
Additionally GaN is typically considered the ``strongest'' substrate for RE with systems demonstrated across 2 layers of graphene. 
GaN's charge decay distance, however, is less that SiC at all layers of graphene, even when RE has only been reported for SiC only up to one graphene layer. 

We conclude that examining the charge density decay from the surface of a material is not a viable metric for determining if it will host RE.
These results also highlight the importance of including explicit layers of graphene within the modeling simulation, which was not always done in early RE work.
The charge density decay distance changes significantly as graphene is included and a realistic accounting for RE viability needs to include the change in bonding the graphene induces. 
In the next section, we demonstrate how this consideration of graphene also needs to extend to the reconstruction that the inclusion of graphene may induce. 

\subsection{Reconstruction Matters}
\begin{figure*}
\includegraphics[width=\textwidth]{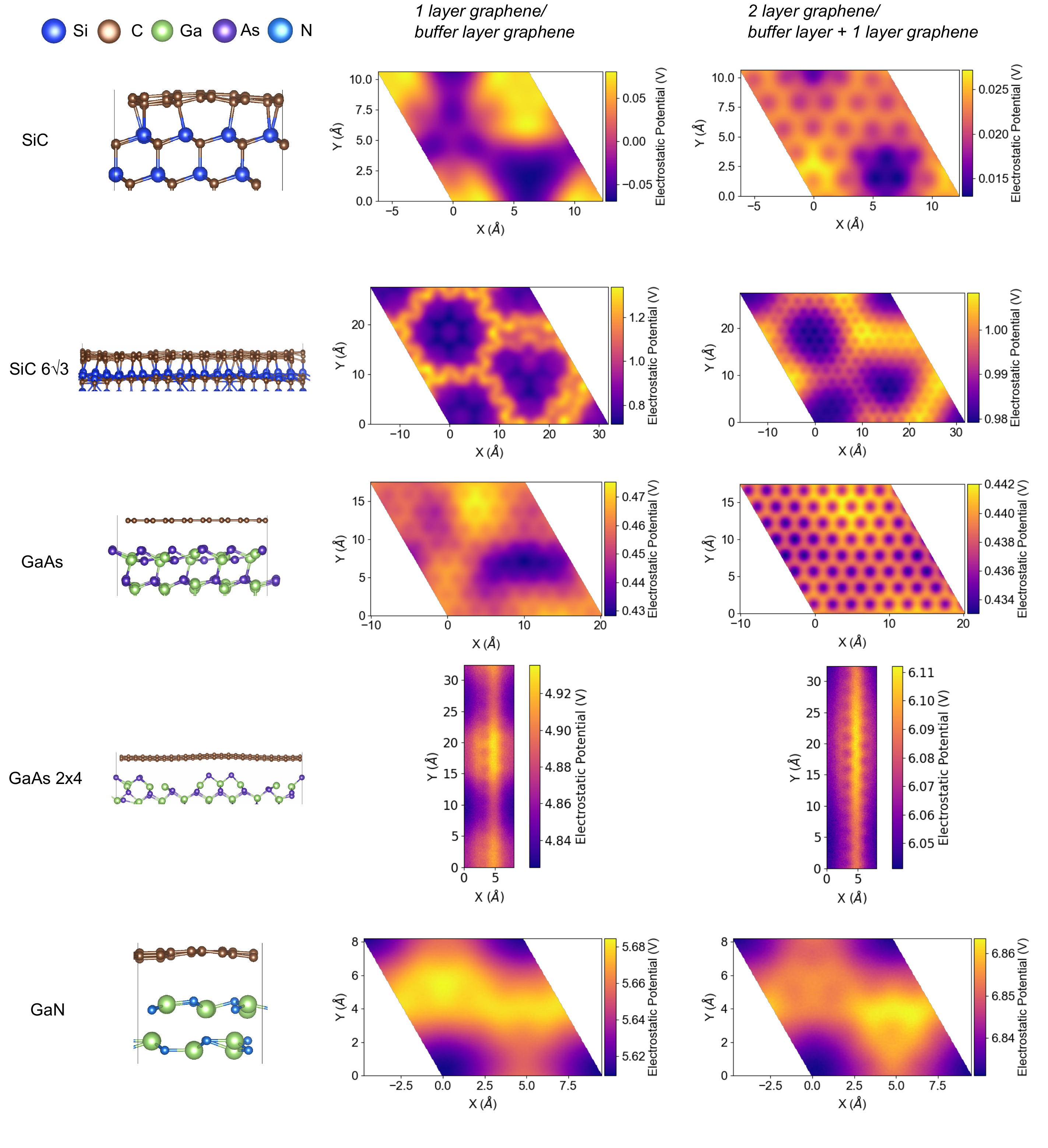}
\caption{ The electrostatic potentials 3 \AA\ above the surface of RE growth substrates. We show the electrostatic potentials for SiC with no reconstruction, the SiC $6\sqrt{3}\times6\sqrt{3}$ reconstruction, GaAs with no reconstruction, the GaAs (100) 2$\times$4 reconstruction, and GaN. We show the electrostatic potentials for when the substrate is covered with one and two layers of graphene. For the SiC case, the first layer of graphene bonds with the underlying silicon atoms and is often called the buffer layer graphene. We show this in the same column as one graphene layer results for the other materials,  and a graphene layer on top of the buffer layer as being equivalent to the two graphene layer cases in other systems. 
\label{fig:reconstruct}}
\end{figure*}

The most common metric that has been used by recent literature to explain remote epitaxy is the electrostatic potential above the substrate. 
In this telling, more polar substrates will have electrostatic potentials that extend further above the surface, through multiple layers of graphene \cite{kong2018polarity}. 
These electrostatic potentials would then influence how atoms which are coming down from epitaxy would bond with the substrate. 
Presumably, there is then an ideal range of electrostatic potentials where the bond with the surface is strong enough to create a defined lattice (\textit{i.e.} not vDW epitaxy), but weak enough that strict lattice matching requirements no longer apply and defect densities can be reduced (\textit{i.e.} not direct epitaxy). 
Previous work \cite{laduca2025transparent} has already identified a number of problems with the simplistic analysis, including that there is often a somewhat arbitrarily chosen background subtraction.

In this work, we directly examine how the electrostatic potential of different surfaces correspond with RE. 
In Fig.~\ref{fig:reconstruct}, we show the electrostatic potential, as calculated with DFT, 3 \AA\ above the surface  for multiple substrates and layers of graphene.
When the graphene is not fully planar as in the SiC case, we calculate the average position of the carbon atoms and take the electrostatic potential 3 \AA\ above that average position.

The clearest takeaway is that surface reconstructions significantly impact the resulting electrostatic potentials.
This is most evident when looking at SiC. 
The widely reported 6H-SiC (0001) $6\sqrt{3}\times 6\sqrt{3}$R30 reconstruction \cite{owman1996sic} leads to much higher absolute electrostatic potentials, and even to much higher differences between the maximum and minimum electrostatic potential for a given surface. 
It also results in a larger hexagonal pattern in the electrostatic potential, which would influence any growth from the resulting substrate. 
Similarly the GaAs (100) 2$\times$ 4 reconstruction \cite{biegelsen1990surface} changes the pattern and symmetry of the electrostatic potentials, resulting in a periodic peak and trough of the electrostatic potential in the $y$ direction. 
Any analysis of RE based on electrostatic potential needs to include these reconstructions for a serious picture of what the growth environment looks like. 
\begin{figure}
    \centering
    \includegraphics[width=0.5\linewidth]{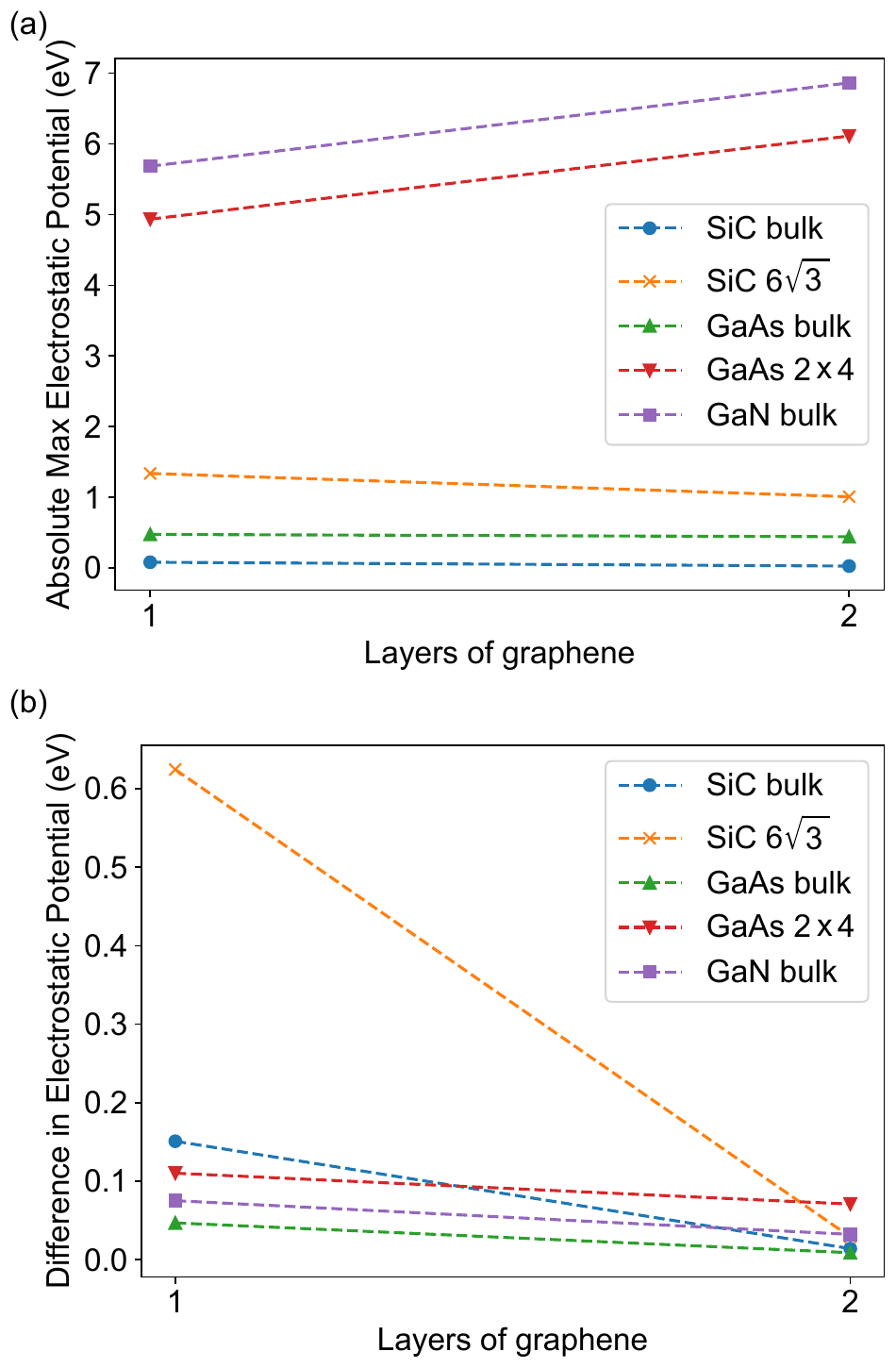}
    \caption{(a) The max electrostatic potential, as shown in Fig.~\ref{fig:electrostatic-pot} for each substrate and (b) the largest difference in electrostatic potential for each of the different substrates.}
    \label{fig:electrostatic-pot}
\end{figure}

Accepting the need for including reconstructions, what information can we infer about how well a RE process will proceed from the electrostatic potential?
In Fig.~\ref{fig:electrostatic-pot}a, we plot the maximum electrostatic potential seen for each system as a function of the number of layers of graphene. 
Surprisingly, this number increases for the GaAs 2$\times$4 reconstruction as well as for GaN as the number of graphene layers increases. 
This alone rules out the maximum electrostatic potential as a viable metric for defining RE, as the potential is expected to decrease as a function of the number of graphene layers, and RE has been shown not to work in GaAs systems through two layers of graphene \cite{kim2017remote}.
A more plausible story relies on understanding the difference in electrostatic potential between the maximum and minimum point.
Based on this, incoming atoms or molecules would be drawn to different locations on the surface with stronger potentials, leading to ordering. 
The preference for these stronger potential sites, however, would be based on the relative difference between this potential and the weaker potential sitees.
In Fig.~\ref{fig:electrostatic-pot}b, we show how the difference between the maximum and minimum electrostatic potential changes for different substrates as a function of the number of graphene layers.
The SiC $6\sqrt{3}$ reconstruction immediately stands out for having a strong electrostatic potential difference with one layer of graphene.
This can be attributed to the bonding that occurs between this first layer of graphene and substrate, often called the buffer graphene layer.
Recent work by Jung \textit{et al.} \cite{jung2025remote} has explored growth on the SiC buffer layer graphene in more detail, examining the difficulty of distinguishing between remote epitaxy and direct epitaxy on a reconstructed surface. 
For these systems, the difference in potential does reliably decrease as the number of graphene layers increases, making it a better as a measure of RE.
There does not exist a clear threshold, however, at which RE is viable. 
The GaAs 2$\times$4 system with two layers of graphene has a larger electrostatic potential difference than the GaN system, which contradicts experimental expectations for RE.
Furthermore, the SiC $6\sqrt{3}$ with two graphene layers has almost the same electrostatic potential difference as the GaN system, again largely against experimental expectations.  

We conclude that while electrostatic potentials can be suggestive, they do not fully explain which materials are RE active.
They do conclusively demonstrate, however, the importance of including realistic reconstructions in any attempt to explain epitaxial processes. 
In the next section we explore the importance of including interaction between the substrate and film elements for describing RE.

\subsection{The Film Matters}

How remote epitaxy depends on the interaction between a substrate and film has been somewhat obscured by a significant fraction of the literature focusing on remote homoepitaxy- growing a film of the same material as the substrate. 
Studies looking at differing film and substrate materials, however, have shown that remote epitaxy is viable for some combinations (e.g. GaAs/graphene/GaAs), but not others with the same substrate (e.g. Ge/graphene/GaAs).
The viability of remote epitaxy therefore clearly depends, to some extent, on the bonding interaction between film atoms and the underlying substrate atoms. %I'm going to slightly hesistate on calling this bonding interaction 
A minority of the theoretical studies have previously examined differing substrate and films, with Wang \textit{et al.} noting a decreasing binding energy for full layers of film as a function of the number of graphene layers \cite{wang2023modulation}.
It is unclear, however, when and how this interaction would take place during the growth process.

\begin{figure*}
\includegraphics[width=\textwidth]{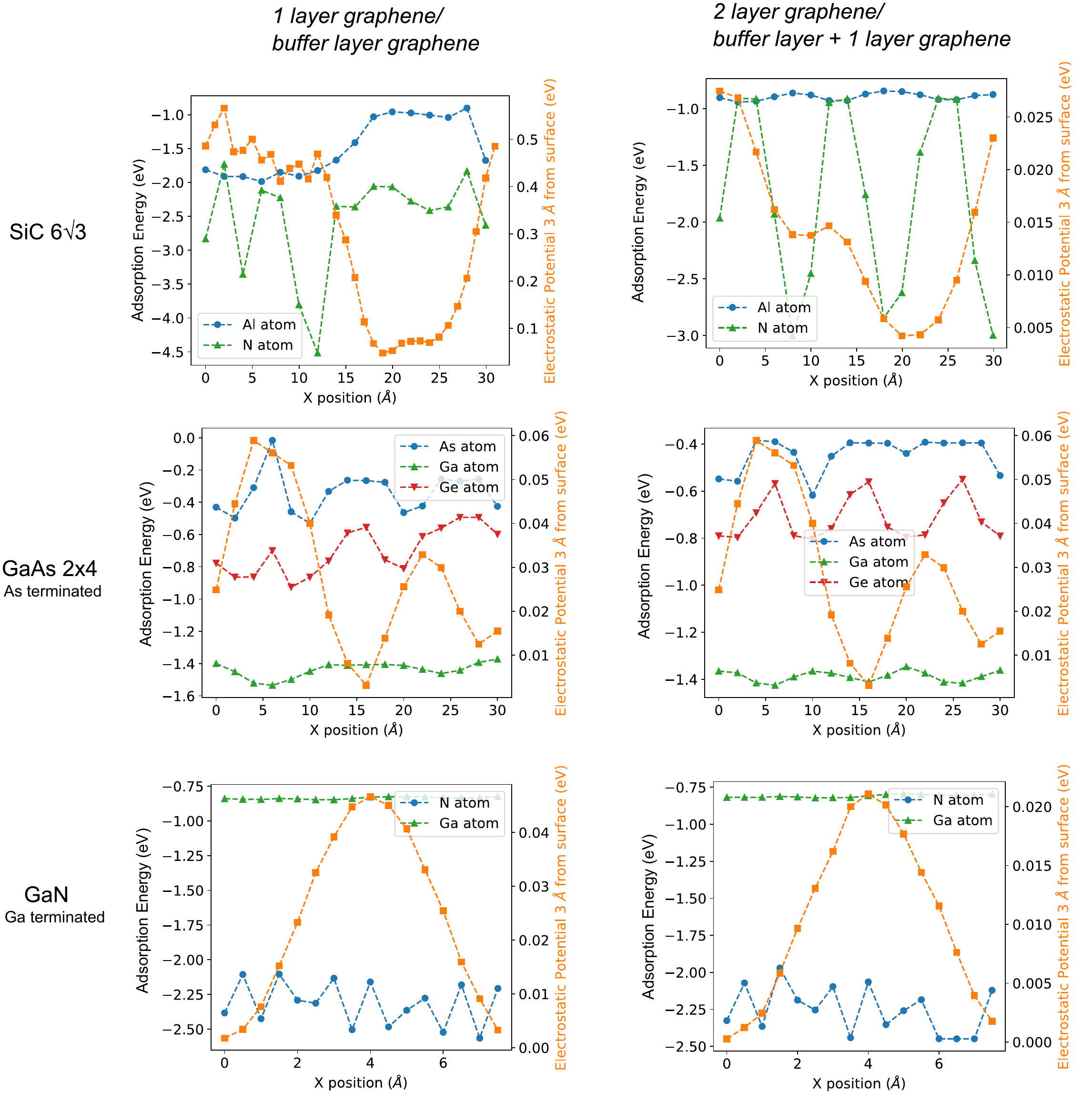}
\caption{ 
\label{fig:ind-atoms} The adsorption energy of individual atoms above the surface of SiC, GaAs, and GaN, with one or two layers of graphene included above the surface. 
The corresponding electrostatic potential is plotted in orange squares for each system.}
\end{figure*}

One plausible hypothesis is that as individual atoms or molecules adsorb to the graphene surface, they adsorb more strongly at certain sites corresponding to the underlying lattice. 
This would then form the basis of the film forming a single crystalline layer; the adsorption sites are determined by the underlying substrates interactions, creating a unified pattern.
To test this hypothesis, we placed individual atoms of film material above substrates with differing layers of graphene, with the resulting adsorption energies shown in Fig.~\ref{fig:ind-atoms}.
In our convention, a more negative adsorption energy indicates that the atom is more strongly bonded to the surface. 
To avoid all the atoms drifting to the strongest potential sites, we fix the position of the atom in the lateral x and y dimensions and only allow each atom to move closer or further from the interface, finding the optimal bonding distance.

The first takeaway is that the adsorption energies do strongly depend on the specific atom of the film tested.
In SiC, for example, N atoms tend to have lower adsorption energies than Al atoms.
These adsorption energies can also be significantly spiky.
While the adsorption energy for N atom on a buffer layer of graphene on SiC is usually on the order of -2.0 eV, it can occasionally get as low as -4.5 eV at specific locations. 
This indicates a strong spatial preference when exposed to the SiC buffer layer graphene surface. 
Similarly, N adsorption energies vary wildly even further away from the SiC substrate with an epitaxial layer of graphene above a buffer layer graphene.
Since this trend extends beyond just one layer of graphene, it is unlikely to be a key indicator of remote epitaxy.
In GaAs with an As terminated surface, the atom with the strongest adsorption energy is Ga, matching expectations of continuing the underlying structure of the substrate.
This is true for both one layer and two layers of graphene on top of GaAs, and the magnitude of the adsorption energies stays roughly the same in both cases. 
Ge atoms also have notably stronger adsorption energies than As atoms, demonstrating that the preference for continuing the ordering of layers through graphene layers is strong.
Finally, in GaN with a Ga terminated surface, we see a much stronger adsorption energy for N atoms above the graphene surfaces.
Notably the adsorption energy for these atoms does not follow any clear pattern or preference for a specific location. 
This is particularly surprising as GaN is considered the most polar material and we might expect strong localization of the adsorption energy above the Ga atoms in the substrate.

In all of these plots, we also include the electrostatic potential at the corresponding locations. 
It becomes clear that electrostatic potential above the surface is not associated with any pattern in the adsorption energy of atoms.
This is most obvious looking at the GaN systems, where the electrostatic potential is forming a clear sinusoidal wave and the adsorption energies for Ga are essentially flat, and the stronger adsorption energies of the N atoms are significantly noisy. 
This further emphasizes that electrostatic potential above the surface does not necessarily provide strong indications of how resulting remote epitaxy growths will take place. 

\begin{figure}
    \centering
    \includegraphics[width=0.5\linewidth]{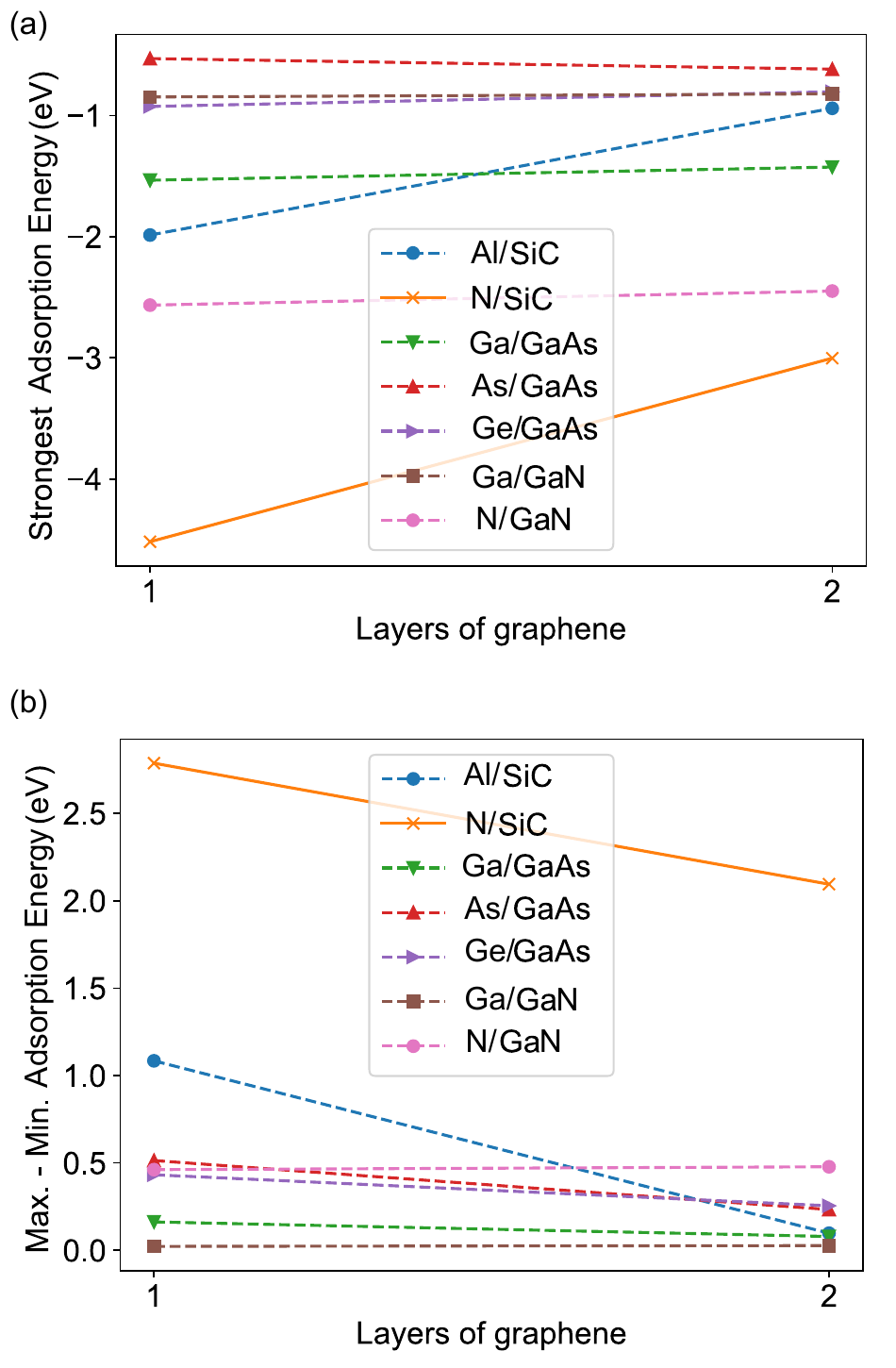}
    \caption{(a) The strongest adsorption energy for each atom above a system and (b) the difference between the maximum and minimum adsorption energy, both as a function of the layers of graphene. 
    Here, the notation on the legend of Al\textbackslash SiC indicates the trend applies to the Al atoms on the SiC substrate.}
    \label{fig:adsorption-e-corr}
\end{figure}

We next test the hypothesis that individual atoms adsorption onto the surface determines remote epitaxial success in Fig.~\ref{fig:adsorption-e-corr}.
We first examine the strongest adsorption energy for a given atom above a susbstrate. 
This would be an indication that strongly bonding a single atom in a fixed location provides a key seed for the growth of a resulting layer.
For the majority of systems, however, there does not appear to be a clear tapering off of adsorption energy with increasing layers of graphene. 
This is particularly contradictory in the case of GaAs, as experimental evidence shows that remote epitaxy should not be expected beyond one layer of graphene. 
We therefore conclude that a single adsorption site is not likely to be determinative for remote epitaxy. 

A more plausible explanation, however, is that atoms on the surface of a graphene covered substrate can diffuse around the surface. 
Systems with stronger differences between their maximum and minimum adsorption energy would have higher barriers for surface diffusion and thus behave more systematically like traditional epitaxy. 
Systems with smaller differences between their maximum and minimum adsorption energy would have high surface diffusion and behave similarly to van der Waals epitaxy. 
Therefore remote epitaxy systems would likely have an intermediate level of adsorption energy differences. 
We test this in Fig.~\ref{fig:adsorption-e-corr}b. 
The evidence is more in favor of this hypothesis, with the difference in adsorption energies decreasing for all systems as the number of graphene layers increase. 
The barrier for N movement in SiC systems still seems high at over 2.0 eV, even with two layers of graphene. 
The counter example to this trend however, is the Ge atom on GaAs having the same adsorption energy range as As on GaAs, both of which have significantly higher differences than Ga atoms on GaAs, which is nominally what the next layer should be. 
This contradicts experimental evidence demonstrating remote epitaxy of GaAs on GaAs, but not Ge on GaAs.

We therefore conclude that the energetics of single atoms of the film are not sufficient to explain remote epitaxy.
Instead, a more complicated picture involving the growth process and kinetics of film formation is necessary.

\subsection{Kinetics Matters}

After single atoms/molecules are deposited on substrate surfaces, they often agglomerate into islands which can move around the surface until full layers are formed. 
The exact movement of these islands is highly system dependent, but is thought to be a key feature in alleviating stress and reducing defect formation. 
Systems with lower sliding barriers, the energetic barrier for an island moving across the surface, tend to have lower defect concentrations since islands can easily move to optimal locations for creating a complete surface layers \cite{matthews1974defects}.
Systems with higher sliding barriers, in contrast, are more likely to have higher defect concentrations and grain boundaries as these islands are not able to easily move into ``optimal'' locations. 
Indeed, previous work has shown that the inclusion of a graphene layer does significantly reduce the migration barrier of a surface film compared to diffusion across a bare substrate \cite{bae2020graphene}.
It is reasonable to hypothesize then, that remote epitaxial systems must have sufficiently low sliding barriers that surface islands can diffuse around the surface, alleviating stress.
If the sliding barrier is too low, however, then the influence of the substrate on the system would be minimal and the growth would resemble van der Waals epitaxy with no clear orientation.
Remote epitaxial sliding barriers should then exist in a narrow band-- strong enough that the underlying lattice periodicity still applies, but weak enough that relaxation of defect densities can still occur.

\begin{figure*}
\includegraphics[width=\textwidth]{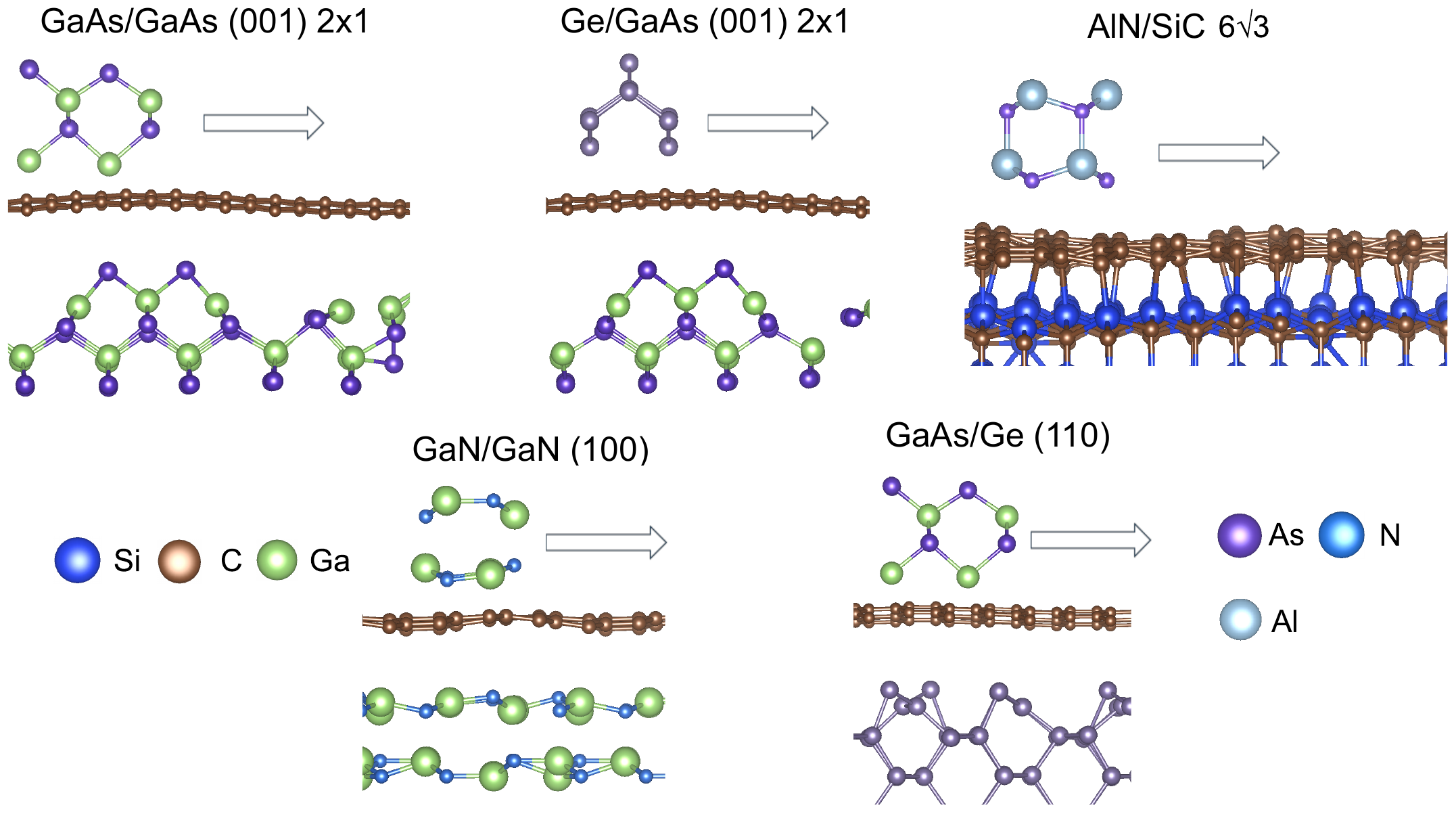}
\caption{ Atomic structure diagrams of the different island/substrate combinations simulated. In these systems, we simulate moving the island across the surface in 1 \AA\ increments. In each island simulation, the atomic positions are allowed to fully optimize. 
\label{fig:islands}}
\end{figure*}

To test this hypothesis, we next simulate islands of film material above multiple substrates with differing layers of covering graphene, as shown in Fig.~\ref{fig:islands}. 
We then move the islands across the surface allowing all atomic positions to relax. 
We measure the energy at each location and determine the sliding barrier as described in the methodology.
To normalize the results, we then divide the sliding barrier by the exposed surface area of the island in question.
We report the sliding barrier results of these calculations in Table \ref{tab:island-sliding-bar}.

\begin{table}[]
\centering
\begin{adjustbox}{width=\textwidth}
\begin{tabular}{llllll}
                 & GaAs/GaAs (100) 2x4 & Ge/GaAs (100) 2x4 & AlN/SiC $6\sqrt{3
                 }$& GaN/GaN (001) & GaAs/Ge (110) \\
1 Layer Graphene & \cellcolor{LightGreen} 0.055               & \cellcolor{LightCoral} 0.007              & \cellcolor{LightGreen}0.020       & \cellcolor{LightGreen}0.097         & \cellcolor{LightCoral} 0.010         \\
2 Layer Graphene & \cellcolor{LightCoral} 0.004               &                    & \cellcolor{LightCoral} 0.008       & \cellcolor{LightGreen}0.015        &               \\
3 Layer Graphene &                     &                    &             &\cellcolor{LightCoral} 0.006        &              
\end{tabular}
\caption{\label{tab:island-sliding-bar} The Sliding Barrier of islands on surfaces with different amount of graphene layers. Energy is reported in terms of eV/\AA$^2$. We shade cells green which have barriers $>$0.01 eV/\AA$^2$ and cells with barriers $\leq$0.01 eV/\AA$^2$ red. This shading notably corresponds with which film/substrate pairs have been experimentally demonstrated for remote epitaxy as shown in Fig.~\ref{fig:intro}.}
\end{adjustbox}
\end{table}

The island sliding barrier results correlate with experimental evidence of systems where remote epitaxy has been observed.
None of the sliding barriers feature values above 0.1 eV/\AA$^2$, which would likely tip the scales of the system closer to direct epitaxy with all its respective defects.
However, there is a clear delineation around 0.01 eV/\AA$^2$. 
Systems with sliding barriers higher than this value have been demonstrated to be effective for remote epitaxy; systems with sliding barriers lower than this value have all not demonstrated success for remote epitaxy in experiment, showing multi-phase data in resulting EBSD characterization. 
This implies that the motion of islands on the surface of graphene is, in fact, \textit{the} defining factor that distinguishes remote epitaxy.
The intervening graphene layers allow islands to slide across the surface smoother than if these were directly on the substrate.
They do not, however, allow the island to move totally freely. 
In the cases where more graphene layers are added to the system, such as three graphene layers on GaN, and two graphene layers above GaAs, the sliding barrier reduces significantly, allowing many phases to form and van der Waals epitaxy to result. 

Why do these island sliding barriers correlate with remote epitaxy when the similar metric of the maximum difference in adsorption energy on individual atoms did not correlate? 
We hypothesize that the film needs a significant surface area on which the underlying lattice potentials can be felt. 
Individual atom motion is too dominated by individual large spikes in the adsorption energy, island motion averages out this fluctuation. 
Another intriguing question relates to the exact quantitative value of the threshold between van der Waals and remote epitaxy. 
We chose 0.01  eV/\AA$^2$ empirically by examining how this sliding barrier metric corresponds to experimentally characterized systems.
It does, however, imply a clear kinetic limit that would be useful to explore in future modeling of growth. 
Another clear question is what size of island is needed for successful RE?
The islands we simulate are typically on the order of $<$ 100 \AA$^2$. 
This is still quite small in the scheme of a full layer growth, but at what point does the system cross over from behaving as individual atoms to islands with distinct sliding barriers? 
When these islands agglomerate, do the same barriers apply?
Further kinetic modeling is needed to fully answer all of these intriguing mechanistic questions. 

\section{Summary and Future Work}\label{sec:conclusions}
So what determines whether remote epitaxy between an arbitrary film and substrate will be successful?
And how many graphene layers can be added before turning into van der Waals epitaxy?
In this paper, we develop a clear metric for predicting remote epitaxy success from first-principles simulations: the sliding barrier of an island of film on the substrate/graphene surface. 
This metric combines information from a number of physical mechanisms that have previously been proposed: the larger surface area of an island allows it to respond to the underlying polarity and electrostatic potential of the substrate.
Measuring the sliding barrier for a specific island of material ensures that the interaction between the film and substrate is explicitly accounted for. 
The success of the island sliding barriers metric implies that the kinetics of initial film formation are key to the success of the remote (and likely any) epitaxy process. 

Throughout this work, we did not include any of the material complications which we know to be significant challenges for experimental systems; e.g. pinholes in the graphene as well as step edges in the system. 
Future simulations work will incorporate these features into the larger sliding barrier calculations. 
The success of sliding barriers as a first-principles metric, however, implies that these graphene defects are not necessarily needed to explain all the demonstrated remote epitaxy systems.
Indeed, Jung \textit{et al.} have recently demonstrated a GdAuGe growth on SiC with a single buffer graphene layer which shows different rotation than both GdAuGe directly on SiC and on multiple epi layers of graphene \cite{jung2025remote}.
Kim \textit{et al.} have also recently shown a clear difference in pinhole mediated epitaxy leading to pyramids whereas clean remote epitaxy resulting in uniform films, again suggesting a large kinetic component to the growth process \cite{kim2025roles}; \textit{i.e.}, remote epitaxy takes place when it is easier for stress in islands when developing to relax instead of forming new surface facets and grain boundaries as in typical epitaxy.
These results together imply that remote epitaxial cannot be solely attributed to pinholes in the graphene.
Future simulations will also work to build a larger kinetic picture of the growth of an entire film. 
This paper demonstrates that atomic level insights can be used to predict novel remote epitaxy pairs, and should likely serve as the basis for future kinetic models. 

\begin{acknowledgement}\label{sec:acknowledgement}
    
    This work was supported by the by the Laboratory Directed Research and Development (LDRD) program at Sandia National Laboratories under project 233271.
    This work was performed, in part, at the Center for Integrated Nanotechnologies, an Office of Science User Facility operated for the U.S. Department of Energy (DOE) Office of Science.
    Sandia National Laboratories is a multi-mission laboratory managed and operated by National Technology \& Engineering Solutions of Sandia, LLC (NTESS), a wholly owned subsidiary of Honeywell International Inc., for the U.S. Department of Energy’s National Nuclear Security Administration (DOE/NNSA) under contract DE-NA0003525. This written work is authored by an employee of NTESS. The employee, not NTESS, owns the right, title and interest in and to the written work and is responsible for its contents. Any subjective views or opinions that might be expressed in the written work do not necessarily represent the views of the U.S. Government. The publisher acknowledges that the U.S. Government retains a non-exclusive, paid-up, irrevocable, world-wide license to publish or reproduce the published form of this written work or allow others to do so, for U.S. Government purposes. The DOE will provide public access to results of federally sponsored research in accordance with the DOE Public Access Plan.
\end{acknowledgement}

\bibliography{refs}

\end{document}